\documentclass{article}

\usepackage{amssymb, mathrsfs, amsthm} 
\usepackage[centertags]{amsmath}
\usepackage[utf8]{inputenc}
\usepackage[T1]{fontenc}
\usepackage{color}
\usepackage{graphicx}
\usepackage{subfigure}
\usepackage[noblocks,affil-sl]{authblk}             
\usepackage{booktabs,graphicx,epstopdf,bbm}
\usepackage{floatrow}
\usepackage{csquotes}
\usepackage[margin=1.25in]{geometry}

\theoremstyle{plain}
\newtheorem{thm}{Theorem}[section]

\newtheorem{proposition}[thm]{Proposition}

\newtheorem{definition}[thm]{Definition}

\newtheorem{remark}[thm]{Remark}

\newcommand\de{\mathrm{d}}
\newcommand\BM{Brownian Motion}
\newcommand\BS{Black-Scholes}

\newcommand\real{\mathbb{R}}
\newcommand\ind{\mathbbm{1}}

\newcommand\LT{local time}
\newcommand\media{\mathbb{E}}

\makeatother
\begin{document}

\title{Estimating the Counterparty Risk Exposure by using the  Brownian motion local time}

\author[1]{Michele BONOLLO \thanks{michele.bonollo@imtlucca.it,mbonollo@numerix.com}}
\author[2]{Luca DI PERSIO \thanks{luca.dipersio@univr.it}}
\author[3]{Luca MAMMI \thanks{luca.mammi@unicredit.eu}}
\author[4]{Immacolata OLIVA \thanks{immacolata.oliva@univr.it}}


\affil[1]{IMT Lucca, Iason Ltd and Numerix LLC}
\affil[2]{Dept. of Computer Science, University of Verona}
\affil[3]{Unicredit Group}
\affil[4]{Dept. of Economics, University of Verona}


\maketitle

\begin{abstract}
In recent years, the counterparty credit risk measure, namely the default risk in {\it Over The Counter} 
(OTC) derivatives contracts,  has received great attention by banking regulators, specifically within the 
frameworks of  \emph{Basel II} and \emph{Basel III}. More explicitly, to obtain the related risk figures, 
one has first obliged to compute  intermediate output functionals related to the \emph{Mark-to-Market} (MtM) 
position at a given time $ t \in [0, T]$,  $T$ being  a positive, and finite, time horizon. The latter implies 
an enormous  amount of computational effort is needed, with related highly time consuming  procedures to be 
carried out, turning out into significant  costs. To overcome latter issue, we propose  a smart exploitation 
of  the properties of the (local) time spent by the Brownian motion close to a given value.
\end{abstract}

\section{Introduction} \label{intro}
For some years now, due to the occurrence of events leading to the financial crisis between 2007 and 
2008, regulators have forced financial institutions to adopt \emph{ad-hoc} procedures	 to 
predict, and therefore prevent, defaults. In other words, banks have to be able to measure and manage their 
default risk. As concerns both the credit and the counterparty risk, in 2006 the Basel Committee for Banking Supervision 
has inserted in the well known \emph{Basel II} reform,  two rather general  methodologies for calculating banks 
capital requirements, namely: the \emph{Standardized approach} and the 
\emph{Internal approach}. While the former one is based on the use of ratings from External Credit Rating Agencies,the latter envisages the evaluation of  certain risk parameters, such as the Exposure at Default (EAD), \cite{BCBS2006}. 

An interesting perspective concerns the so-called \emph{Counterparty Credit Risk,} (CCR), 
which represents  the default risk linked to {\it Over The counter} (OTC) derivatives contracts. The latter case implies the computation, as intermediate outputs, of a large set of different functionals related to the \emph{Mark-to-Market,} ($MtM$) of the position over a future time horizon, at a given time  $ t \in [0,T]$, where $T<+\infty$ is the {\it time horizon}.  Standard techniques for the evaluation of such an exposure are based on classical Monte Carlo  methods, which are characterized by a strong dependence on the number of considered assets and related high  computational time costs, see, e.g., \cite{Liu}. 

Other approaches have been also given, as an example the considering geometric point of view, see \cite{ST}, or general ambit stochastic processes as in \cite{DPP}, 
or some optimal investment control problems, as in \cite{ChevLySc13}, 
even if, as a general benchmark, the Monte Carlo set of methods are the widest used.
Nevertheless, as mentioned, Monte Carlo techniques are far from being computationally satisfactory, even in simple cases.
For example, a medium bank requires \textbf{$D = O(10^4)$} derivative deals and 
\textbf{$U = O(10^3)$} risk factors, evaluated in \textbf{$K = 20$} time steps with \textbf{$N = 2000$} 
simulations, which allow for \textbf{$K \cdot N \cdot U = 4 \cdot 10^7$} grid points for the risk factor simulation and 
\textbf{$K \cdot N \cdot D = 4 \cdot 10^8$} tasks for deals evaluation.

To overcome latter drawbacks, the literature has recently proposed new techniques, e.g., the 
Vector Quantization \cite{vq,cfg1,cfg2,CaSa}, or more enhanced hardware technologies, such in the case of \emph{grid computing} and 
\emph{Graphics Processing Units} (GPUs) capabilities, see, e.g., \cite{Castagna},\cite{PW} and references therein. 
In the context of American option pricing, other methods recently investigated are the martingale-based approach \emph{\`a la Rogers,} 
see e.g. \cite{Lelong}, and the simple least-squares approach, see \cite{AntIssMech15},\cite{glass} for further details.  
A different solution can be achieved exploiting the so called {\it polynomial chaos expansion} approach, see, e.g., \cite{BeSco},
\cite{DiPPel} and references therein. 

Another possi\-bi\-li\-ty consists in exploiting the properties of suitable mathematical tools, as for the case of 
derivatives pricing via \emph{Brownian \LT.}
Given a probability space $(\Omega, \mathfrak{F},\mathbb{P}),$ we consider a standard  \BM\, $\{W_t\}_{t \geq 0}$ defined on it. 
Then, for $\omega \,\in\, \Omega$ and a 
level $a,$ an interesting point is to determine how much time the sample path $W_t(\omega)$ spends 
close to $a$.
A possible answer dates back to the works  written by
Paul L\'evy in 1948, where the author introduced the concept of \emph{Brownian Local Time,} see \cite{levy}.  


The right approach consists in defining the Brownian Local Time, BLT from now on, as the following density: 
\begin{equation}\label{def1_LT}
L_t(a) := \frac{1}{2\epsilon}\lim_{\epsilon\longrightarrow 0}\mu\{x : |x-a|\leq \epsilon\} \;, 
\end{equation}

where $\mu$ represents the Lebesgue measure on the real line.
\begin{remark}{}
It is worth to mention that there does 
not exist a standard notation to define the BLT, since some authors prefer to multiply the limit 
in Eq. ~\eqref{def1_LT} by $\frac{1}{4\epsilon},$ instead of by $\frac{1}{2\epsilon},$ see, e.g., \cite{KS}.
\end{remark}
More formally, the \LT\, can be defined through the so-called \emph{occupation formula}, see \cite{KS}, namely by the following equation
\begin{equation} \label{occ_form}
\int_0^t f(W_s) \de s = 2 \int_{\real} f(x) L_t(x) \de x \;,
\end{equation}
where the left-hand side is a random measure, called \emph{occupation measure} or 
\emph{sojourn measure,} at fixed time $t$ and level $x \,\in\,\real$, while $f$ is a $L^1$ function, 
$f:\, \real \,\rightarrow\, \real.$ We refer to Section ~\ref{loc_time} for a more detailed 
discussion of the BLT properties.
To what concerns the fine properties of the \LT, e.g., the identification of both its distribution function and related 
density function and moments, we refer to \cite{DY}, \cite{KS}, \cite{tak}, and references therein.
It is also worth to mention that
there exists an extended literature dealing with the theoretical applications of BLT such as 
an extension of It\^{o}'s formula to convex functions, the definition of the density of the occupation 
measure for a \BM\, with respect to the Lebesgue measure, see e.g. \cite{vq}, etc. 

On the other hand, a relatively 
limited literature has been devoted to concrete applications of BLT and its properties. 
The latter lack can be easily recognized in  frameworks related to Economy and Finance. Nevertheless theoretical aspects of the 
BLT can be fruitfully exploited to analyze a wide range of financial tools, particularly with respect 
to the pricing of some kinds of exotic path-dependent options as in the case, e.g., of    \emph{range accrual options} and  
\emph{accumulators,} where the payoff depends on the time spent by the underlying below or above a given  level, resp. between two boundaries, resp. outside of them, see e.g. \cite{Mij}. 
Moreover, the use of BLT is almost absent in the risk management field.
The present work aims at filling latter gap by showing that the numerical
integration of the BLT density function can be used to  evaluate the risk exposure, hence obtaining results 
that are very compelling when compared with classical Monte Carlo benchmark algorithms. 

The paper is organized as follows: in Section ~\ref{CCR_fr} we introduce the financial framework, focusing on the 
regulatory viewpoint, and with  emphasis to the instructions for calculating the EAD and the Credit Value Adjustment (CVA), then, in Section ~\ref{mathset}, the mathematical setting is introduced also recalling 
the main properties of the BLT while, in section ~\ref{proposal}, we provide
 the {\it local time approach} to the aforementioned type of financial problems, also analyzing  its performances compared to more standard 
techniques with respect to  an EAD application; eventually, in Section ~\ref{concl}, we state our main conclusions and we outline future line of research.

\section{Counterparty Risk: the Financial Framework} \label{CCR_fr}

\subsection{The Credit Counterparty Risk in the Basel approach} \label{CCR_basel}

In the Basel II framework, the \emph{ Counterparty Credit Risk,} CCR from now on, is a specific class of the 
broader credit risk category. Let us recall the definition of the Basel committee, shortly BCBS, as it is written in \cite{BCBS2006}: 
\begin{definition}{}\label{BCBSdef}
\emph{Counterparty Credit Risk (CCR) is the risk that the counterparty in a trans\-action could default before the 
final settlement of the transaction's cash flows. An economic loss would occur if the transactions or portfolio of 
transactions with the counterparty has a positive economic value at the time of default.}
\end{definition}

Unlike a firm's exposure to credit risk through a loan, 
CCR creates a bilateral risk of loss: the market value of the transaction is uncertain, it can be positive 
or negative to either counterparty and can vary over time with the movement of underlying market factors. 
A typical example is given by IRS. 
Several classes of financial transactions are considered in the regulatory perimeter, but most of the CCR arise 
from \emph{Over the Counter} (OTC) derivatives, in the peer-to-peer relationships with a defaultable counterparty. 
From a practical perspective, the buyer of any option, or the holder of a derivative with positive $MtM$, both are facing a 
CCR. If the two counterparties agree upon a \emph{netting set,}, e.g. a running compensation process in their deals, 
the current exposure will be given by the positive part of the algebraic sum of all deals.

As in the whole Basel setting, the risk must be dealt with by setting apart an amount regulatory capital 
of the bank which is linked to the risk measure  called \emph{capital requirement}, let us indicate it by $K$, and specified  in \cite{BCBS2011}, as follows
 
\begin{align} \label{cr}
    K & = EAD \cdot 1.06 \cdot LGD \left\{\Phi\left[\left(\frac {1}{1-\rho}\right)^{0.5} \Phi^{-1}(PD) 
		+ \left(\frac {\rho}{1-\rho}\right)^{0.5} \Phi^{-1}(0.999)\right]-PD \right\}\cdot c \;,
\end{align}
where:
\begin{itemize}
	\item 
	$EAD$ is the \emph{Exposure at Default,} namely an estimation of the extent to which a bank may be 
				exposed to a counterparty in case of default;
	
	\item 
	$LGD$ is the \emph{Loss Given Default,} namely an estimation of the percentage of the credit not 
				recoverable in case of insolvency;
	
	\item 
	$PD$ is the \emph{Probability of Default,} namely an estimate of the likelihood that a default will occur; 
	
	\item 
	$\rho$ is the \emph{asset return correlation coefficient};
	
	\item 
	$c$ is a constant which takes into account some maturity adjustment and can vary with respect to different regulatory 
				\emph{portfolios,} such as enterprise or retail loans;
	
	\item 
	$1.06$ is a coefficient depending on the calibration procedure made by the Basel committee;
	
	\item 
	$\Phi$ is the cumulative distribution function of a standard Gaussian random variable;
	
	\item 
	$\Phi^{-1}$ is simply the inverse of $\Phi$, also referred to as 
				the \emph{quantile function}. 
\end{itemize}


As well highlighted in the BCBS definition, see Def.~\eqref{BCBSdef}, the \emph{EAD} estimation makes the counterparty risk very 
different from the normal credit risk for loans and mortgages. In fact, the Basel formula ~\eqref{cr} requires a $1$ year 
measurement process, and the default time $\tau$ could be, or  it could not to be, in any future time $t.$ 

For a mortgage, we know the future exposure profile, since it can be computed using the amortizing plan. 
Differently, in the CCR, the \emph{EAD} estimation is fairly difficult, because of two different reasons: the future 
exposure is stochastic and, further, it depends on the market parameters via its specific evolution pricing model. 

In other words, the CCR depends both on the credit parameters $(PD,LGD)$ and on the market influenced $EAD$ parameter 
in its magnitude, that is why it is also referred as the \emph{boundary risk.} To summarize, the CCR has to be determined according to eq. ~\eqref{cr} for the credit risk, but its $EAD$ input estimation is itself a hard challenge, to 
which the Basel committee and the financial operators pay most of their  attention.

\subsection{Exposure and CVA calculation in the Basel II-III setting} \label{CVA}
In order to calculate the $EAD$ quantity in the CCR context by a robust and conservative way, the Basel II framework
\cite{BCBS2006} defines two important different approaches: the \emph{Standard model} 
and the \emph{Internal model}, also called \emph{EPE-based} approach. 

In the standard model, we have $EAD = MtM + \mbox{\emph{Add-On}},$ 
where the \emph{Add-On} is computed exploiting a table which depends on both the underlying asset class and on 
the time to maturity. In this case, the idea is that such an \emph{Add-On} takes into account the future volatility 
by additive coefficients. As an example, for an equity option with maturity $M$ years and such that $1 \leq M \leq 5$, we have
that the \emph{Add-On} is $8\%$ of the notional amount, while for an interest rate derivative it is just $0.5\%.$
In the \emph{EPE-based} approach, to which the present work refers, some notation has to be pointed out.

\begin{itemize}
\item Given a derivative maturity time $0<T<+\infty$, we consider $K\in\mathbb{N}^{+}$ time steps 
			$0<t_{1}<t_{2}<\cdots<t_{K},$ which constitute the so called \emph{buckets array}, denoted by 
			${\bf B}^{T,K}$, where usually, but not mandatory, $t_{K}=T.$ 

\item 
For every $t_{k}\in{\bf B}^{T,K}$, we denote by $MtM\left(t_{k},S_{k}\right):=MtM\left(t_{k},S_{t_{k}}\right)$
			the fair value, \emph{Mark-to-Market}, of a derivative at time bucket $t_{k},$ with respect to the 
			\emph{underlying value} $S_{k}$ considered at time $t_{k}$.
			
\item 
For every $t_{k}\in{\bf B}^{T,K}$, we denote by $MtM\left(t_{k},S^{k}\right):=MtM\left(t_{k},S^{t_{k}}\right)$
			the fair value (\emph{Mark-to-Market}) of a derivative at time bucket $t_{k}$, with respect to the whole 
			sample path $S^{k}:=\left\{ S_{t}:0\leq t\leq t_{k}\right\},$ and with initial time $t_{0}=0.$ 
			
\item 
Taking into account previous definitions, we indicate by $\varphi=\varphi\left(T-t_{k},S_{k},\Theta\right)$
			the pricing function for the given derivative, where $\Theta$ represents the set of parameters from which 
			such a pricing function may depend, e.g., the free risk rate $r$ or the volatility $\sigma.$ 
\end{itemize}

We give an account of the main amounts, as they are defined 
in Basel III \cite{BCBS2006},  that will be used later on to estimate the EAD.

We denote the \textit{Expected Exposure} of the derivative at time $t_k \,\in\, {\bf B}^{T,K}$ $(EE_k)$, as follows 
\begin{equation} \label{f_EE}
EE_{k}:=\frac{1}{N}\sum_{n=1}^{N}MtM\left(t_{k},S_{k,n}\right)^{+},\; N\,\in\,\mathbb{N}^{+}\;,
\end{equation}
which is the arithmetic mean of the positive part of the $N$ Monte Carlo simulated
MtM values, computed at the $k-$th time bucket $t_{k},$ with respect
to the underlying $\mathrm{S}.$

\begin{remark}{}
The\emph{ positive part} operator is effective
if we are managing a symmetric derivative, such as an interest rate
swap or a portfolio of derivatives. Nevertheless it is redundant if we consider a single option, 
as the fair value of the option is always positive from the buy side
situation. We want to stress that the sell side does not imply counterparty
risk, hence it is out of context.
\end{remark}

We evaluate the \emph{Expected Positive Exposure} (EPE) as follows 
\begin{equation} \label{f_EPE}
EPE:=\frac{1}{T} \sum_{k=1}^{K}EE_{k}\cdot\Delta_{k},
\end{equation}
where $\Delta_{k}=t_{k}-t_{k-1}$ indicates the time space between
two consecutive time buckets at the $k$-th level. If the time buckets
$t_{k}$ are equally spaced, then the formula reduces to $EPE=\frac{1}{K}{\sum_{k=1}^{K}EE_{k}}$.
Therefore, the EPE value gives the time average of the $EE_{k}$ and reflects the hypothesis that the default 
could happen, as a first approximation, at any time with the same probability. 

We define  the \emph{Effected Expected Exposure} as follows
$$EEE_{1} :=EE_{1}; \mbox{ and } EEE_{k} := \max\left\{ EE_{k},EEE_{k-1}\right\}, \, k=1,\ldots,K \;,$$
observing that, due to its non decreasing property, $EEE_{k}$ takes into account the fact that,
once the time decay effect reduces the MtM as well as the counterparty
risk exposure, the bank applies a roll out with some new deals. 

We also define the \emph{Effected Expected Positive Exposure} (EEPE) by
$$
EEPE:=\frac{1}{T}\sum_{k=1}^{K}EEE_{k}\cdot\Delta_{k}\;.
$$

\begin{remark}{}
In order to avoid too many inessential regulatory details, we will
work on $EE_{k}$ and EPE, the others quantities being just arithmetic modifications
of them.
\end{remark}

In what follows we shall rewrite previously defined quantities in
continuous time, and we add the index $A$ to indicate the \textit{adjusted}
definitions. Moreover we consider the dynamics of the underlying $S_{t}:=\left\{ S_{t}\right\} _{t\in[0,T]}$,
$T\in\mathbb{R}^{+}$ being some expiration date, as an It\^{o} process,
defined on some filtered probability space $\left(\Omega,\mathfrak{F},\mathfrak{F}_{t\in[0,T]},\mathbb{P}\right)$.
As an example, $S_{t}$ is the solution of the stochastic differential
equation defining the geometric Brownian motion,  $\mathfrak{F}_{t\in[0,T]}$
being the natural filtration generated by a standard Brownian motion
$W_{t}=(W_{t})_{t\in[0,T]}$ and with respect to  a complete probability
space $\left(\Omega,\mathfrak{F},\mathbb{P}\right)$, where $\mathbb{P}$ is often referred to as 
the so called \textit{real world} probability measure, or
an equivalent risk neutral measure under the martingale approach to option
pricing, see, e.g., \cite{KS}.

The \emph{Adjusted Expected Exposure} $EE^{A}$ is given by 
\begin{align} \nonumber
EE_{k}^{A} & :=\mathbb{E}_{\mathbb{P}}\left[MtM\left(t_{k},S_{k}\right)^{+}\right] 
= \int\varphi\left(T-t_{k},S_{k},\Theta\right)d\mathbf{\mathbb{P}} \\ \label{eeA}
 & \cong \frac{1}{N}\sum_{n=1}^{N}MtM\left(t_{k},S_{k,n}\right)^{+}=\widehat{EE_{k}^{A}} \;.
\end{align}

Similarly, we define the \emph{Adjusted Expected Positive Exposure} $EPE^{A}$ as follows
\begin{equation}
EPE^{A}:=\int EE_{t}^{A}\de t=\int \int\varphi\left(t,S_{k},\Theta\right)\de\mathbb{P} \de t\;.\label{epeA}
\end{equation}
With respect to latter formulation, the Basel definition is simply one of the
many methods that can be used to estimate the expected fair value of the derivative
in the future.

\begin{remark}{}\label{remark2} We skip any comment about the choice
of the most suitable probability measure $\mathbf{\mathbb{P}}$ to
be used in the calculation of $EE_{k},$ the latter  being beyond the aim
of the present paper. For a detailed discussion on the role played by the \emph{risk neutral}
probability, resp. by the \emph{historical} real world probability, see e.g. \cite{BMP}. 
\end{remark}

\begin{remark}{}\label{remark3} Let us underline that the component usually indicated as {\it discount factor},
or \emph{numeraire}, is missing in the EPE definition, the latter being a byproduct
of the conservative approach used in
the risk regulation. 
\end{remark}

Besides EAD, understood as a CCR measure, also the \emph{Credit Value Adjustment} (CVA) may be specified. 
According to Basel guidelines \cite{BCBS2011}, the $CVA$ represents the capital charge for potential MtM 
losses associated with a deterioration in the credit worthiness of a counterparty. 
Moreover, by introducing the CVA, the expression of the derivative payoff provides a new term, related 
to the value of the security emerging in case of default. 
%
In particular, we have
$$Payoff = \phi(\mathbf{m^T},\mathbf{c})\cdot \ind_{\{\tau > T\}} 
+ RR \phi(\mathbf{m^T},\mathbf{c})\cdot \ind_{\{\tau \leq T\}}$$
where $\tau$ is the counterparty default time, $\phi(\mathbf{m^T},\mathbf{c})$ is the terminal payoff 
at maturity $T$, where $\mathbf{m^T}$, resp. $\mathbf{c}$, stands for the path of the market parameters in $[0,T]$, resp. 
for the contract clauses on which the payoff depends, while $RR := 1 - LGD$ is the so called 
\emph{recovery rate,} that is, the extent to which principal and accrued interest on a debt instrument that is 
in default can be recovered, expressed as a percentage of the instrument's face value.  Hence, the CVA metrics performs an average reduction of the MtM value and involves another form of  risk, the \emph{CVA risk,} characterizing the uncertainty of the future CVA evolution. 
\begin{remark}{}
Let us note that one of the major  credit rating agency, namely {\it Moody's},
estimates defaulted debt recovery rates using market bid prices observed roughly $30$ days after the 
date of default. Recovery rates are measured as the ratio of price to par value, see \cite{moody} for further details.
\end{remark}

\subsection{Computational Challenges}
An interesting and challenging problem consists in the concrete implementation of both the EPE and the CVA. 
Because of the \emph{EPE} (\emph{EAD}) volatility, the counterparty risk must be monitored frequently, hence the 
standard requirement for an internal model validation is a daily frequency.
To have an idea of the magnitude of the computational efforts for such a procedure, let us consider that, in 
a medium size banking group that aims to satisfy the regulators indications, we could observe $D= 10000$ deals 
in the book, $N = 2000$ simulations and $K = 20$ time steps. If we indicate with $PT$ the number of pricing 
tasks for each CCR run, we easily get
\begin{equation} \label{bank}
    PT = D\cdot N \cdot K = 4 \cdot 10^8 \;.
\end{equation}
Latter example easily shows how great  is the required computational effort, even because a large part of the pricing 
algorithms are still represented  by specific {\it \`a la} Monte Carlo techniques. Hence, although the pricing software 
and {\it CPU} features are adequate for front office purposes, they  become unsatisfactory for CCR evaluation 
constraints. Considering the storage requirements, we define the new parameter $\alpha,$ i.e., the number of 
execution cases that have to be stored to allow traceability and auditability of the output results. We can fix 
$\alpha = 13,$ if we suppose an {\it end-of-month} backup with 1 year memory. Of course, the storage is run on different 
record types, e.g., deal information, payoff information, simulation information, etc. 
For the sake of simplicity, we can think about the storage as a unique large record type, let us indicate it by RT, which takes into account all the relevant 
information, hence obtaining
\begin{equation} \label{RT}
    RT = D\cdot N \cdot K \cdot \alpha = 5.2 \cdot 10^9 \;.
\end{equation}
As each {\it record} could easily require $1000$ bytes, hence we raise to $5.2$ terabytes of storage.
In other words, the CCR computation involves the computational hard challenges related to the credit and market 
risk fields. In particular, the high frequency of monitoring implies a number of concrete practical implementation of 
efficient and robust CCR calculation. 
In order to address previous challenges, important results have been achieved exploiting techniques related to the so called 
\emph{BigData} analysis as well as using graphical processing units (\emph{GPU}), see, e.g., the numerical investigations provided in \cite{Castagna}, \cite{PW}. 
Nevertheless, the solution to the computational challenges posed by the CCR evaluation are neither completely, nor 
satisfactory solved by the aforementioned software improvements. That is why there is a growing and wide interest in finding 
more effective theoretical techniques, and related applied algorithmic procedures. 

\begin{remark}{}
We would like to underline that while the Basel Committee generally defines frameworks and principles,  it does 
not prescribe a mandatory model or some numerical technique that one has to apply. Hence, starting from the next section, we propose a novel method   
to perform  the EPE calculation, in the broad CCR setting, by exploiting a BLT approach. 
\end{remark}

\section{Mathematical setting} \label{mathset}
\subsection{The Black-Scholes Market Model} \label{BSmodel}
In what follows we will refer to the celebrated Black and Scholes diffusion process, see \cite{BS}, as a theoretical 
benchmark for our proposal's verification.
Let us consider a financial market, composed by a risk-less security $B,$ with constant return $r,$ and a risky 
asset $S,$ defined by means of a geometric Brownian motion, namely

\begin{equation} \label{market}
\begin{cases}
\de B_t & = r B_t \de t \\ 
\de S_t & = S_t \mu \de t + S_t \sigma \de W_t \\ 
\end{cases}
\;,
\end{equation} 

where $\mu \,\in\, \real, \; \sigma > 0$ and $\{W_t\}_{t \geq 0}$ represents a standard Brownian motion.

The SDE representing the geometric Brownian motion in eq. ~\eqref{market} admits the following unique solution 
\begin{equation} \label{BS_eq}
S_t = S_0 \exp\left\{\left(\mu-\frac{\sigma^2}{2}\right) t + \sigma W_t \right\} \,, 
\end{equation}
which characterizes the dynamic of the underlying of a \emph{derivative,} namely of  a financial instrument that gives 
to its owner a terminal \emph{payoff} $\phi = \phi(\mathbf{m^T},\mathbf{c})$ evaluated at the maturity $T.$ 
As to give an example, in the simple case represented by considering a \emph{European call option,} we have 
$\phi := (S_T - K)^+,$, where the {\it level} $K$ is called the {\it strike price} of the option, since it provides a positive profit 
if and only if $S_T > K.$ 
Let us recall that the parameters $r$ and $\sigma$ represent the \emph{risk-free rate} and the \emph{volatility} of the 
underlying, respectively. The risk-free rate plays a key role in the evaluation process, that is, the definition 
of the fair value, \emph{FV} from now on, at time $0.$ 
In other words, by an application of the It\^{o}-D\"oblin Lemma, it is possible to show that, in the 
fair value evaluation, the actual drift $\mu,$ with $\mu > r,$ and the unknown risk aversion of the market, or 
\emph{utility function,} both disappear, while the fair value can be simply calculated as the discounted expected payoff, 
where the risk-neutral drift $r$ can straightly replace the expected drift $\mu$ in Eq. ~\eqref{BS_eq},   
see, e.g.,  \cite{Hull} for further details. In the basic \BS\, simplified model, where the risk-free rate $r$ is deterministic and constant over time, latter
principle leads to a general evaluation strategy given by

$$FV_t = E[e^{-r(T-t)} \phi(\mathbf{m^T},\mathbf{c_t})] \;.$$

The \BS\, model received several extensions and criticism, e.g. : 
sophistication in the \emph{payoff algebra,} due to the natural innovation process in the financial markets. They allow 
to cover the effective requirements or to get new profits by issuing new appealing products. 

Generally speaking, we can have several clauses, e.g., bundling of different strikes, barriers, memory effects, 
occupation time clauses, etc.c,  or the dependence of $\phi$ on the whole sample path of $S_t,$ as it happens when dealing with th so called \emph{Asian, 
look-back} options; 
 new models for the \emph{underlying,} that arise from the different dynamics among the asset classes, e.g., considering interest 
rates {\it versus} equity {\it versus} forex,  or from the need of a better calibration of the empirical data, e.g., volatility surface {\it versus} 
flat volatility. As a benchmark model for an interest rate underlying, the \emph{Vasiceck model}, see \cite{Vas}, and the 
\emph{Hull-White model}, see \cite{HW}, usually replace the \BS\ model; 
an increase in the number of \emph{risk sources,} e.g., by taking into account the stochastic behavior of 
volatility, as it happens in the \emph{Heston model}, see \cite{Heston}. 
For a complete review of models, resp. of pricing formulas, see \cite{Hull}, resp. \cite{Haug}. 

Nevertheless, let us recall that, if dealing with a whole portfolio of financial instruments, independently from their 
features, the Mark-to-Market dynamics, 
can be adequately fitted by a log-normal process because of the 
compensation or aggregation effect among se\-ve\-ral single position returns. This is a common practice in the asset 
management sector, often referred to as  the \emph{normal portfolio} approach, see, e.g., \cite{saita}. 

Moreover, also in the risk ma\-na\-ge\-ment approach, the lognormal \BS\, model is quite satisfactory, as pointed out, e.g., in
\cite{glass} where a particular type of \emph{incremental risk charge} (IRC) model has been proposed.
We recall that, in the real world, one buys or sells a derivative for a given quantity, or {\it notional}, namely  one 
\emph{takes a position.} Hence, in the following, we will often replace the fair value by its related  \emph{Mark-to-Market} 
expression $(MtM),$ hence by the fair value equipped with a quantity and a sign.
\subsection{Local time and occupation time} \label{loc_time}
Let $\{W_t\}_{t \geq 0}$ be a standard \BM\, defined over the probability space $(\Omega,\mathfrak{F},\mathbb{P}).$ 
The Local Time for the \BM\, $W_t,$, or equivalently the \emph{Brownian Local Time} (BLT), first  introduced by P. L\'evy in \cite{levy}, can be seen 
as a stochastic process indicating the amount of time spent by the Brownian motion process close to a given {\it level}  $a \,\in\, \real.$ 
To quantify such a random time, in \cite{levy} the author  introduced the following random field

$$L_t(a) = \frac{1}{2\varepsilon} \lim_{\varepsilon \,\rightarrow\, 0} \mu\{0 \leq s \leq t, \,:\, 
|W_s - a| \leq \varepsilon \} \;,$$ 

where $t \,\in\, [0,T], \, a \,\in\, \real$ and $\mu$ is the Lebesgue measure. $L_t(a)$ was defined as the \emph{mesure de voisinage}, and L\'evy proved its existence, 
its finiteness and its continuity, see \cite{levy}. 
More rigorously, let us recall the following useful definition

\begin{definition}{}
The random field  $\{L_t(x,\omega) :  (t,x) \in [0,T] \times \real,\omega \in \Omega \}$
is called a \emph{Brownian Local Time} if the random variable	$L_t(x)$ is $\mathfrak{F}$-measurable, 
the function $(t,x) \,\longmapsto\, L_t(x, \omega)$ results to be continuous and 
\begin{equation}\label{occupation}
\Gamma_t(B,\omega) := \int_0^t \ind_B (W_s) \de s =  \int_B  L_t(x,\omega) \de x \;,
\end{equation}
with $0 \leq t \leq \infty$ and $B \,\in\, \mathcal{B}(\real)$.
\end{definition} 
Let us also recall that
the quantity in the left-hand side of ~\eqref{occupation} is known as the \emph{occupation time} of the Brownian motion  up to time $t.$
A crucial theoretical point consists in establishing the BLT existence. This is ensured by the 
\emph{Trotter Existence Theorem},  see, e.g.,  
\cite[Thm 6.1.1, Ch. 3]{KS} for details. The Brownian Local Time satisfies several useful properties.For the sake of convenience, we report only the ones that we are going to use for our computational purposes, while we refer the interested reader to \cite[Section 3.6]{KS}, for a more comprehensive treatment of the subject as well as for the proofs of the results which we will exploit in what follows.

\begin{proposition}{} \label{prop}
For every Borel-measurable function $f \,:\, \real \,\rightarrow\, [0,T],$ we have 
\begin{equation}\label{occ_form2}
\int_0^t f(W_s(\omega)) \de s = \int_{\real} f(x) L_t(x,\omega) \de x \;, 0 \leq t \leq T \;.
\end{equation}
\end{proposition}

As a consequence of Prop. ~\eqref{prop}, we have 

\begin{equation} \label{occ_form3}
\int_0^t \ind_{\real}(W_s) \de s = \int_{\real} L_t(x, \omega) \de x = t \;.
\end{equation}

The following Proposition is known in the literature as the \emph{Tanaka-Meyer decomposition,} 
see \cite{KS} for further details. 

\begin{proposition}{} \label{TanakaMeyer}
Let us assume that the BLT exists and let $a \,\in\, \real,$ be a given number. Then, the process $\{L_t(a)\}_{0 
\leq t \leq T}$ is a nonnegative, continuous, additive functional which satisfies 
\begin{equation}\label{TM}
L_t(a) = (W_t - a)^+ - (z - a)^+ - \int_0^t \ind_{(a,+\infty)}(W_s) \de W_s \;,
\end{equation}
for $0 \leq t \leq T$ and for every $z \,\in\, \real.$
\end{proposition}

\begin{remark}{}
It is worth to mention that the representation given in Prop. ~\ref{TanakaMeyer}, can be generalized to a semimartingale. 
\end{remark}

The \BM\, spends a random time over any set $A,$ hence it is important to be able to derive its density, namely, the 
probability that the BLT stands close to a given {\it level} $a$, for a time $\de y.$ Such a  density is given by
\begin{equation} \label{LT_bm}
g(y;t,a) = \sqrt{\frac{2}{\pi t}} e^{-\frac{(y + |a|)^2}{2t}} \;,  
\end{equation}
see \cite[Eq. 1.3.4, p. 155]{BoSa}. 

\section{The Local Time Proposal for the CCR} \label{proposal} 

\subsection{An application of Brownian Local Time in finance: the Accumulator Derivatives} \label{acc}

In what follows we focus our attention on a particular type of derivatives, namely the \emph{Accumulator}, which is 
a  path-dependent forward enhancement without a guaranteed 
worst case. More precisely, an \emph{Accumulator} is characterized by a  contract, agreed upon two parties, 
which provides that the investor purchases/sells a pre-determined quantity of stock at a settled strike price $K,$ on 
specified observation days $t_1, \ldots, t_n, \, t_n \leq T,$ $T$ being the expiry of the contract. 
Usually, an \emph{Accumulator} is linked to an underlying which is an exchange rate, but we have 
similar payoffs with different names, \emph{range accrual}, in the broad interest rate derivatives frameworks.
An example is given by the \emph{FTSE Income Accumulator,} identified through the ISIN code \emph{XS1000869211,} 
over the \emph{FTSE 100 Index,} with plan start date on February, 14th 2014, plan end date on August, 14th 2020, 
and maturity date on August, 28th 2020. 
The plan is expected to pay every three months, the level depending on how the FTSE 100 Index has performed over 
the quarter. The maximum income is $6.75\%$ every year, paid if the underlying closes between $5000$ and $8000$ 
points on each weekly observation date. Otherwise, the income will proportionally be reduced, according to the 
time spent out of the range. 

Although such a kind of derivative product exhibits some benefits, e.g., a noticeable improvement of the 
exchange rate, the lack of product costs and the existence of several tailor-made features, on the other hand 
there are some drawbacks. 
The latter allowed the accumulator derivatives to earn the nickname of \emph{``I will kill you later''} products. 

In order to permit more flexibility and to reduce hedging costs, the accumulator contracts may include one or 
two \emph{knock-out barriers,} in order to restrict the maximum profit and/or the maximum loss by the investor. 
Basically, if at the end of the $i$-th observation day, the closing price $S_i$ of the underlying hits 
the barrier $H,$ for all $i = 1, \ldots, n,$ then the option stops. 

We distinguish among \emph{accumulator-out one-sided knock-out, accumulator-in one-sided knock-out, 
accumulator-out range knock-out, accumulator-in range knock-out,} depending on whether the investor 
purchases (resp., sells) a one-sided or range knock-out call (resp., put) and sells (resp., purchases) 
a one-sided or range knock-out put (resp., call), with the same strike price, fixing dates and expiry date. 
Hence, the payoff $\mathscr{P}_i$ of an accumulator derivative at the observation day $t_i, \, i = 1, \ldots, n,$ is given by 
\begin{equation} \label{payoff}
\mathscr{P}_i = 
\begin{cases} 
0, & \mbox{if } \max\limits_{0 \leq \tau \leq t_i} S_{\tau} \geq H \\ 
Q(S_{t_i} - K), & \mbox{if } \max\limits_{0 \leq \tau \leq t_i} S_{\tau} < H, \; S_{\tau} \geq K \\ 
g Q (S_{t_i} - K), & \mbox{if } \max\limits_{0 \leq \tau \leq t_i} S_{\tau} < H, \; S_{\tau} < K \\ 
\end{cases} 
,
\end{equation} 

where $Q$ is the purchase quantity and $g$ is the gearing ratio, both fixed by contract, see, e.g., 
\cite{AccPrice} for further details. 
For our purposes, we set $Q = 1$ and $g = 2$, hence implying that the fair value $FV$ is given by
\begin{equation} \label{FV_acc}
FV_i = \sum_{j=1}^N \left[C_{t_j} - P_{t_j}\right] \cdot e^{-r(T-t_i)} 
\end{equation}
where $C_{t_j}:=C(S_0,K,T-t_j,\sigma,H)$, resp. $P_{t_j}:=P(S_0,K,T-t_j,\sigma,H)$, represents the fair price of a knock-out call 
option, resp. of knock-out put one. We recall that, by assuming that the underlying evolves according to the \BS\, 
model,  the call price and the put price appearing in Eq. ~\eqref{FV_acc} have a closed form, see, e.g., \cite{AccPrice}. 
\subsection{The Proposal for the \emph{EE} evaluation} \label{prop_EE} 
In what follows we show how the \LT\ may be used as a handy tool in the evaluation of the 
Counterparty Credit Risk (CCR) for accumulator derivatives. 
%

In the setting described by eqs. ~\eqref{market}-~\eqref{BS_eq}, 
it is still possible to determine how long the geometric \BM\, remains in the neighborhood of 
any point $a,$ for  any given set. In other words, we could attain the density of local time with respect to a 
geometric \BM, see, e.g.,  \cite{BoSa} for further details. In particular, we have
\begin{align} \label{LT_gbm}
\nonumber
P & (L(t,a) \,\in\, \de y) = f(y;t,a,\sigma,\nu,S) 
= \sqrt{\frac{2}{\pi t}} \sigma a \left(\frac{a}{S}\right)^{\nu} 
e^{-\nu^2 \sigma^2 \frac{t}{2} - \frac{(\sigma^2 a y + |\log(a/S)|)^2}{2 \sigma^2 t}} \\ \nonumber
& + |\nu|\sigma^2 a \left(\frac{a}{S}\right)^{\nu} \left[ e^{-|\nu|(\sigma^2 a y + |\log(a/S)|)} 
Erfc\left(\frac{\sigma^2 a y + |\log(a/S)|}{\sigma\sqrt{2t}} - |\nu| \sigma 
\sqrt{\frac{t}{2}}\right) \right. \\ 
& \left. - e^{|\nu|(\sigma^2 a y + |\log(a/S)|)} 
Erfc\left(\frac{\sigma^2 a y + |\log(a/S)|}{\sigma\sqrt{2t}} + |\nu| \sigma \sqrt{\frac{t}{2}}\right) \right] \;, 
\end{align}

where $t$ represents the time up to which the \LT\ is evaluated, $a$ is the underlying, $\sigma$ is the 
volatility parameter, $\nu := -\frac{1}{2} + \frac{r}{\sigma^2}$, $r$ being the risk-free rate, $S$ represents the 
spot price, and $Erfc(z)$ is the complementary error function, namely
$$Erfc(z) = 1 - Erf(z) \;,  Erf(z) = \frac{2}{\sqrt{\pi}} \int_0^z e^{-x^2} \de x \;.$$ 

For ease of convenience, from now on we will not consider the presence of knock-out barrier. 

By recalling the expressions of the payoff and the fair value stated in eqs. ~\eqref{payoff} and ~\eqref{FV_acc}, and  
supposing a high fixing frequency, we obtain
\begin{align} 
\nonumber
\mathcal{P}^{(LT)} & = \sum_{j=1}^N [(S_{t_j} - K)^+ - 2 (K - S_{t_j})^+] \\ \nonumber
 & \approx \int_0^T [(S_{t} - K)^+ - 2 (K - S_{t})^+] \de t \\ \label{payoff_LT}
 & = \int_{\real} L(T,x) [(x - K)^+ - 2 (K - x)^+] \de x \;,
\end{align} 
where the last equality in eq. ~\eqref{payoff_LT} follows exploiting  eq. ~\eqref{occ_form2}, while $L(t,x)$ is the 
BLT  up to maturity $T.$ As a consequence, we are able to  evaluate the corresponding fair value for every observation day 
$t_i, \, i = 1, \ldots, n,$
\begin{align} 
\nonumber
FV_{t_i}^{(LT)} & = e^{-r(T-t_i)} \media \left( \int_{\real} L(T,x) [(x - K)^+ 
- 2 (K - x)^+] \de x \right) \\ \nonumber
 & = e^{-r(T-t_i)} \int_{\real} \media[L(T,x)] \left[(x - K)^+ 
- 2 (K - x)^+\right] \de x \\\label{FV_LT}
 & = e^{-r (T-t_i)} \int_{\real} \int_0^{\infty} y f(y;T,x,\sigma,\nu,S) 
\left[(x - K)^+ - 2 (K - x)^+\right] \de y \de x  \;,
\end{align}
basically as an application of the Fubini Theorem, in the second equality, and by the very definition of the BLT density given in eq. ~\eqref{LT_gbm}. 
Hence, as an intermediate first application, we use the above pricing formula for our {\it Accumulator}, and we compare 
three different pricing techniques for the {\it Accumulator} defined by $(C - 2P),$ where $C$ and $P$ are the Call option 
price and the Put option price, 
namely: BSD, the straight BS evaluation, i.e. Eq. ~\eqref{FV_acc};  BSC, the continuous time version of BSD, described in Section ~\ref{num_res}; 
LT: the time proposal given by formula ~\eqref{FV_LT}.
For a more detailed discussion of the aforementioned quantities, i.e. concerning BSD,BSC and LT, see the Section ~\ref{num_res}.
The results have been reported in the table below and they have been obtained setting $S_0 = 1$, with $N = 250$ fixing dates.
We can see that the accuracy is very good, with just a small decay when the volatility
parameter increases.
\begin{table*}[!ht]
\centering
\caption{{\footnotesize{Comparison among fair values obtained with the three methods.}}}
\label{tab_FV}
\begin{tabular}{|c|c|c|c|c|c|c|}
\hline
\textbf{r}	&	\textbf{K}	&	$\boldsymbol{\sigma}$	&	\textbf{FV BSD}	&	\textbf{FV BSC}	
&	\textbf{FV LT}	&	$\boldsymbol{\Delta(LT,BSD)}$ \\
\hline
0,01	&	0,9	&	15\%	&	0,0961	&	0,0961	&	0,0961	&	0,00\% \\
0,01	&	0,9	&	25\%	&	0,0783	&	0,0784	&	0,0781	&	-0,26\% \\
0,01	&	1	&	15\%	&	-0,0323	&	-0,0322	&	-0,0322	&	-0,31\% \\
0,01	&	1	&	25\%	&	-0,0587	&	-0,0585	&	-0,0576	&	-1,87\% \\
0,02	&	0,9	&	15\%	&	0,1008	&	0,1008	&	0,1008	&	0,00\% \\
0,02	&	0,9	&	25\%	&	0,0837	&	0,0837	&	0,0839	&	0,24\% \\
0,02	&	1	&	15\%	&	-0,0248	&	-0,0247	&	-0,0247	&	-0,40\% \\
0,02	&	1	&	25\%	&	-0,0509	&	-0,0508	&	-0,0501	&	-1,57\% \\
\hline
\end{tabular}
\end{table*}
We are interested in evaluating $EE$ and $EPE$ introduced in Section ~\ref{CVA}, hence, for all $t_i, \, i = 1, \ldots, n,$ we have
\begin{align} \label{EE_LT}
EE_{t_i}^{(LT)} & = \media\left(FV_{t_i}^{(LT)}\right) \\ 
EPE_{t_i}^{(LT)} & = \frac{1}{T} \int_0^T \int_{\real} e^{-r(T-t)} \media(L(T,x)) [(x - K)^+ - 2 (K - x)^+] \de x \de t \;.
\end{align}

\begin{remark}{}
By recalling that the expectation functional $\media$ imply an integration task, it results that $EPE$ requires 
the evaluation of  a triple integral. So that,  we have two further integration steps with respect to the 
usual $MtM$ current evaluation of the deal, against the expectation with respect to the market parameters 
scenarios and the time average, respectively. 
\end{remark}
\begin{remark}{}
We wonder which probability measure is better to use, when the expectation functional is evaluated. In other terms, 
we are interested in choosing the most appropriate distribution at any time $t$ and for all market parameters, which 
represent the input data for the pricing function. 
As it is well-known in the literature, there are two alternatives, namely the risk neutral distribution, 
or the historical one. 
Since we mainly focus on computation issues, we believe that the latter is not a relevant point. Anyway, 
in agreement with the majority of the authors,  we 
follow the convention of adopting the  historical distribution.
In the \BS\, framework, the latter implies that there is a real world drift $\mu$ different from the risk-free rate $r,$ 
and such that $\mu > r.$
\end{remark}
\subsection{Application and Numerical Results} \label{num_res}
To the extent of testing the goodness of the our \LT\ proposal to estimate the $EE$ as well as the$EPE,$ 
we compare the algorithm described in the previous Subsection with a benchmark \emph{ \`a la} Black and 
Scholes (BSD). 
First of all, let us fix the number of simulations, indicating them by $Nsim.$ Then, for every simulation, 
	\begin{itemize}
		\item we consider $Nday = 250$ business days, indicated by $t_i, \; i = 1, \ldots, Nday,$ for each of which $t_i,$ we simulate  
			the price of the underlying by using the following {\it discretization procedure}
					\begin{align*}
					S_{t_i} & = S_{t_{i-1}} e^{\left\{\left(r-\frac{\sigma^2}{2}\right)\Delta t_i 
				 + \sigma \sqrt{\Delta t_i} \cdot \mathcal{N}(0,1) \right\}} \;, 
					\end{align*}
					where $\Delta t_i=t_i-t_{i-1} = \frac{1}{Nday} \, ,\, \forall i=1, \ldots, Nday.$
		\item then we compute the {\it Accumulator} price by means of the following formula
					\begin{align} \label{FVBSD}
					FV_{t_i}^{(BSD)} & = \Delta \left(\sum_{j=1}^{i} [(S_j - K)^+ - 2(K - S_j)^+] 
					+ \sum_{k=i+1}^{Nday} [C_{t_k} - 2 \cdot P_{t_k}] \right) \cdot e^{-r(T-t_i)} \;, 
					\end{align}
					where $\Delta := \frac{T}{Nday}$ and $C_{t_k} = C(S_{0},K,r,\sigma,T-t_k),\,P_{t_k} = P(S_{0},K,r,\sigma,T-t_k)$ 
					are the call, resp. the put, price.  
\end{itemize}
In order to evaluate the Counterparty Credit Risk, we choose $10$ time steps, one every $25$ business days, 
then we determine the Expected Exposure $EE^{(BSD)}$ and the Expected Positive Exposure $EPE^{(BSD)}$ by using 
eq. ~\eqref{f_EE}, resp eq.~\eqref{f_EPE}. 
	Since an {\it Accumulator} derivative is characterized by a daily, at least, fixing frequency, then we 
	could take into account a continuous version of the derivative fair value, hence we consider 
\begin{equation}\label{FVBSC}
FV_t^{(BSC)} \cong \int_0^T e^{-r(T-t)} (C_t - 2 P_t) \de t \;,
\end{equation} 
where $C_{t} = C(S_{0},K,r,\sigma,T-t),\,P_{t} = P(S_{0},K,r,\sigma,T-t)$ are the call, resp. the put, price, computed as before.
Eq. ~\eqref{FVBSC} allows us to consider a continuous version of the benchmark, let us indicate it by BSC. In order to 
compare the LT and the BSC approaches, we carry out a time discretization  approximating the BSC by retracing the steps 
of the BSD algorithm and by considering $10^4$ fixing dates, namely $40$ observations per day, instead of one. 
Finally, we are able to appraise the Expected Exposure $EE^{(BSC)}$, resp. the  Expected Positive Exposure 
$EPE^{(BSC)}$, again by exploiting eq. ~\eqref{f_EE}, resp. eq. ~\eqref{f_EPE}.
As regards the \LT\ algorithm, we use a numerical integration, and, in order to have such an integration 
as efficient as possible, we fixed convenient lower and upper bounds. 

\paragraph{Numerical results} To show how the \LT\ techniques behave compared to classical 
approaches, we provide the results reported in Table ~\ref{tab_EPE} which contains the EPE values obtained with methods introduced in the 
previous Sections. 
More precisely, we run all the algorithms for several strike, volatility and risk-free parameters, according 
to the following choices: spot price $S_0 = 5.7$ ; strike price: $K = [4.78, \, 3.75, \, 2.98 ]$; volatility: $\sigma = [0.15, \, 0.2, \, 0.3]$
risk-free rate: $r=[0.01, \, 0.02]$ 
%
%
%
%
We have analyzed the aforementioned three methods, whose values are described in columns 2,3 and 4, focusing on the changes 
$(\Delta)$ of EPE values in columns 5,6,7. Every row is characterized by a triplet $(K_i, \sigma_j, r_h), 
\,i=1,\ldots,3, \; j=1,\ldots,3, \; h=1,2,$ to specify which values of strike price, volatility and 
risk-free rate we refer. 

\begin{table*}[ht]
\centering
\caption{{\footnotesize{Expected Positive Exposure of an accumulator derivative.}}}
\label{tab_EPE}
{\scalebox{0.87}{
\begin{tabular}{|c|c|c|c|c|c|c|}
\hline
$\boldsymbol{(K,\sigma,r)}$	&	\textbf{BSD}	&	\textbf{BSC}	&	\textbf{LT}	&	$\boldsymbol{\Delta(BSD,BSC)}$	
&	$\boldsymbol{\Delta(BSC,LT)}$	&	$\boldsymbol{\Delta(BSD,LT)}$ \\
\hline
$(4.78,0.15,0.01)$	&	0,9303454395	&	0,9303714275	&	0,9303781163	&	-0,00279\%	&	0,00072\%	&	0,00351\% \\
$(4.78,0.2,0.01)$	&	0,9049015095	&	0,9049413280	&	0,9048279190	&	-0,00440\%	&	-0,01253\%	&	-0,00813\% \\
$(4.78,0.3,0.01)$	&	0,8251642939	&	0,8251928714	&	0,8247838254	&	-0,00346\%	&	-0,04957\%	&	-0,04611\% \\
$(3.75,0.15,0.01)$	&	1,9686102762	&	1,9686111645	&	1,9686848833	&	-0,00005\%	&	0,00374\%	&	0,00379\% \\
$(3.75,0.2,0.01)$	&	1,9675941521	&	1,9676004514	&	1,9676336107	&	-0,00032\%	&	0,00169\%	&	0,00201\% \\
$(3.75,0.3,0.01)$	&	1,9547899122	&	1,9548275248	&	1,9547735333	&	-0,00192\%	&	-0,00276\%	&	-0,00084\% \\
$(2.98,0.15,0.01)$	&	2,7348505526	&	2,7348507540	&	2,7349168463	&	-0,00001\%	&	0,00242\%	&	0,00242\% \\
$(2.98,0.2,0.01)$	&	2,7348375084	&	2,7348378657	&	2,7348585689	&	-0,00001\%	&	0,00076\%	&	0,00077\% \\
$(2.98,0.3,0.01)$	&	2,7336498018	&	2,7336570672	&	2,7336545073	&	-0,00027\%	&	-0,00009\%	&	0,00017\% \\
$(4.78,0.15,0.02)$	&	0,9556220367	&	0,9556450742	&	0,9558308683	&	-0,00241\%	&	0,01944\%	&	0,02185\% \\
$(4.78,0.2,0.02)$	&	0,9318141129	&	0,9318494415	&	0,9318254905	&	-0,00379\%	&	-0,00257\%	&	0,00122\% \\
$(4.78,0.3,0.02)$	&	0,8548444479	&	0,8548671968	&	0,8542861489	&	-0,00266\%	&	-0,06797\%	&	-0,06531\% \\
$(3.75,0.15,0.02)$	&	1,9871484085	&	1,9871500602	&	1,9873803563	&	-0,00008\%	&	0,01159\%	&	0,01167\% \\
$(3.75,0.2,0.02)$	&	1,9862637277	&	1,9862700504	&	1,9864841609	&	-0,00032\%	&	0,01078\%	&	0,01110\% \\
$(3.75,0.3,0.02)$	&	1,9743416072	&	1,9743766907	&	1,9744396571	&	-0,00178\%	&	0,00319\%	&	0,00497\% \\
$(2.98,0.15,0.02)$	&	2,7496039372	&	2,7496047313	&	2,7497784936	&	-0,00003\%	&	0,00632\%	&	0,00635\% \\
$(2.98,0.2,0.02)$	&	2,7495932160	&	2,7495941376	&	2,7497625846	&	-0,00003\%	&	0,00613\%	&	0,00616\% \\
$(2.98,0.3,0.02)$	&	2,7485174039	&	2,7485245347	&	2,748661624	&	-0,00026\%	&	0,00499\%	&	0,00525\% \\
\hline 
\end{tabular}}}
\end{table*}

\begin{remark}{}
Let us underline the meaning of the three $\Delta$ comparisons in the right part of the above 
Table. $\Delta(BSC,BSD)$ does not take into account our proposal, but it measures the difference 
between the real world (BSD), where the fixing is discrete over time, and its continuous version, namely the BSC one.
$\Delta(BSC,LT)$ has a double role. On one hand, it measures the rightness of our algorithm implementation, 
as the two methods are theoretically equivalent. Once we verify that the difference is small, with a 
more practical perspective it allows us to monitor the numerical accuracy of the tools we used to 
perform the various numerical integration involved in both the techniques. Finally, $\Delta(BSD,LT)$ 
considers both the previous effects and measures the global accuracy of our BLT proposal, 
where we proxy, by continuous time, the real world problem by a new, local time based, technique. 
\end{remark}

In order to complete the comparison between the different methods proposed, we draw a parallel between 
the execution times of the individual methods, which is reported in Table ~\ref{tab_time}. In particular, 
we invite the reader to dwell on the last two columns, for which the computational effort is comparable. 
We observe that the elapsed time of the \LT\ algorithm is less then the elapsed time 
of the BSD approach and, on average, the former is about half of the latter. 
\begin{table}[!t]
\centering
\caption{{\footnotesize{Average elapsed time of the three algorithms, measured in seconds.}}}
\label{tab_time}
\begin{tabular}{|c|c|c|}
\hline
\textbf{BSC}	&	\textbf{BSD}	&	\textbf{LT} \\
\hline
18.314768 & 4.98831 & 2.12311 \\
\hline 
\end{tabular}
\end{table}

Finally, we exhibit a couple of graphs comparing the errors of the algorithm
LT and BSC, with respect to the \emph{exact} case BSD, once the strike
price and the free risk rate have been set, while the volatility $\sigma$ changes.
We observe that the relative error in very good for small volatilities. To
this extent, we will investigate further the software implementation details.
Anyway, referring to the computational time in the Table ~\ref{tab_time} above,
we think that in the usual trade-off (accuracy,time) the LT approach undoubtedly
dominates the BSC approximation, and it can compete with the
true BSD  model.

\begin{figure}[H]
\centerline{
\resizebox{0.55\textwidth}{!}{\includegraphics{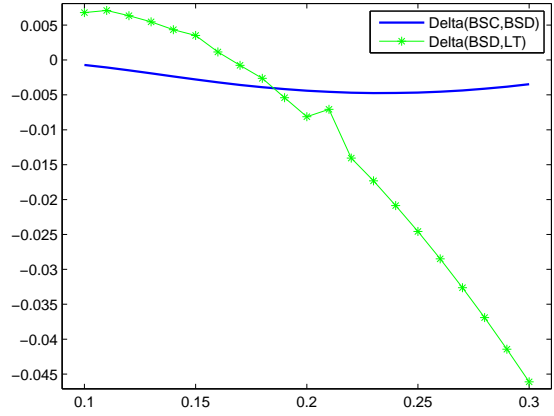}}}
\caption{Comparison between EPE changes for fixed risk-free rates. $\Delta(BSD,BSC)$ vs 
$\Delta(BSD,LT)$ with $r=1\%$ and $K=3.75.$ The volatility parameter varies between 
$10\%$ and $30\%.$}
\label{plot1}
\end{figure}

\begin{figure}[H]
\centerline{
\resizebox{0.55\textwidth}{!}{\includegraphics{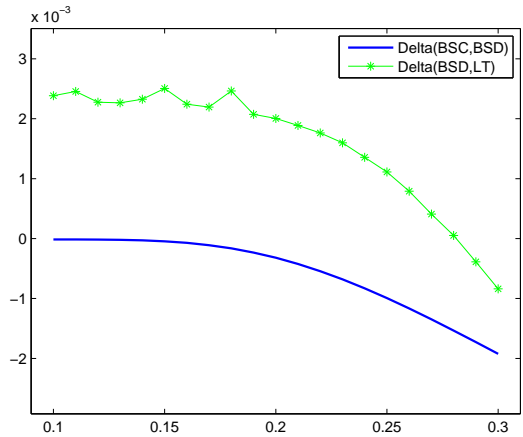}}}
\caption{Comparison between EPE changes for fixed risk-free rates. $\Delta(BSD,BSC)$ vs 
$\Delta(BSD,LT)$ with $r=1\%$ and $K=4.78$ The volatility parameter varies between $10\%$ 
and $30\%.$}
\label{plot2}
\end{figure}

\subsection{Some remarks about computational complexity} \label{comp_compl} 
Once a new methodology or algorithm has been proposed, one would like to picture a 
general analysis of the computation complexity of the new method compared to its more 
traditional competitors; similarly for the accuracy and convergence rate. 
In the simplest and naive case, one has just one parameter, let it be \emph{N}, e.g., 
the number of simulations, the number of deals in the portfolio,
the number of time steps, etc., and the computational complexity
could be stylized by a single ``order'' such as $O\left(N\right),\: O\left(N^{2}\right)$
and so on. 

Despite this elegant theoretical approach, concrete applications are characterised by  
extra difficulties as, e.g.:
(1)
the proposal, or the set of competitors, could depend on some different
	parameters, and $N$ could not be a proper summary of the technique set up;
(2) for each atomic algorithmic task, namely  for any simulation of a loop
			of \emph{N} simulations, the different competitors could contain calculations
			with very different levels of complexity and elapsed time, let them be
			$t_{1}$ and $t_{2}$. Hence it can happen that for small or medium values
			of the parameter \emph{N} the actual computational time of the two
			algorithms do not match the asymptotic order ranking, e.g., it could
			be  that $t_{1}\cdot N>t_{2}\cdot N^{3/2}.$ 
(3) finally, the observed computational times depend on many implementation
			details: the numerical integration method, bounded or not bounded
			integration, efficiency of the  libraries embedded in the
			exploited programming languages.

Coming back  to the above tables of execution times for BSD, BSC and LT, 
 also focusing on the  evaluation of the \emph{EE} and \emph{EPE} values, 
loops behave similarly with cross-method in increasing the number of calculations, and we observe that the BSC involves a 
time integration of the rather complicated BS formula, while the BSD has a complexity given by the 
$BS\cdot t_{n},$, the second term being the number of fixing times, and eventually the LT has a 
complexity given by time-space integration of a quite simple function which is the payoff itself. 

Moreover, we optimized the latter by bounding both the \emph{inf} and the \emph{sup} of the space integral.
Therefore, even without an exhaustive comparison,  also considering
 different  techniques implementations, we can conclude that the LT proposal allows for a good {\it Accuracy versus Effort} trade-off. 
We also  underline that  extensions to other market parameters, clauses and payoffs are needed.

\section{Conclusions and Further Research} \label{concl}
We have addressed the issue of the CCR assessment for the so-called accumulator derivatives, within the \BS\, financial framework 
with one risky asset.   Since the corresponding payoff depends on the time spent by a geometric \BM {\it near} a given 
value, we have exploited the notion of BLT which turns  to play a crucial role in the
derivative pricing step for CCR evaluation. 

However, it is possible to involve BLT also in the risk factors 
simulation step: roughly speaking, for each time bucket $t_k,$ we could employ BLT to build up the grid 
$(t_k, S_{t_{k},n})$ and the corresponding probabilities, and evaluate the $k$-th Expected Exposure $EE_k$ 
as the sum of weighted probability masses. 
We have proposed an original approach
founded on the possibility of 
expressing the BLT in terms of its probability density. 
The associated implementation with regard to EPE evaluation leads to 
numerical results that significantly improve those obtained by
standard procedures \emph{\`a la} \BS. A smaller execution time and a better \emph{EE} appraisal accuracy, 
makes our method a competitive tool,  suggesting to  extend the \LT\ approach to more general 
derivatives, such as barrier options or Asian options.
Next step consists in comparing our results
with those derived 
in \cite{CorDip2014,CorDip2016}.

Moreover, we also plan to use the results presented in \cite{tak}, namely a generalization of the 
well-known L\'evy's \emph{Arc-sine Law}, see also \cite{arcsine2}, which provides the distribution of 
the \emph{occupation time} given in eq. ~\eqref{occupation}. In fact, we intend to use the related Takacs 
formula as an alternative expression for the probability density stated in ~\eqref{LT_gbm} which has been 
extensively used in this paper. 

Finally, we are aware that the one-dimensional case turns out to be unrealistic, though relatively easy to 
implement, albeit to work in the $1$-dimension framework is a very acceptable proxy for derivatives of banks 
with corporate customers, i.e. small and medium size enterprises; in these cases the $i$-th 
customer has a very small number of deals, with main dependence on a single risk factor, e.g. the EUR 
interest rate curve.  After all, a large number of risk factors entails a very hard estimation of correlations. 

To overcome such drawbacks, financial institutions resort to some heuristic and easy-to-extend methods. 
For example, in the $2$-dimensional case it is common practice to consider $\left\langle \de W_t^{(1)}, 
\de W_t^{(2)} \right\rangle = 0$ \emph{between} the asset classes, e.g. interest rate, forex or equity, 
and $\left\langle \de W_t^{(1)}, \de W_t^{(2)} \right\rangle = \de t$ \emph{within} the asset class. 
Such a procedure could be easily extended to the $N$-dimensional case, with $N >> 1.$ This is clearly a 
complicated issue. From a theoretical point of view, the literature provides contributions related 
to the study of multidimensional BLT, see e.g. \cite{BHK2007} and references therein.

\end{document}